\documentclass[conference]{IEEEtran}
\IEEEoverridecommandlockouts
\usepackage{cite}
\usepackage{amsmath,amssymb,amsfonts}
\usepackage{algorithmic}
\usepackage{graphicx}
\usepackage{textcomp}
\usepackage{xcolor}
\def\BibTeX{{\rm B\kern-.05em{\sc i\kern-.025em b}\kern-.08em
    T\kern-.1667em\lower.7ex\hbox{E}\kern-.125emX}}
    
\usepackage{comment}
\usepackage{hyperref}
\usepackage{cleveref}

\usepackage[nonumberlist]{glossaries}
\usepackage{glossary-inline}

\newacronym{CPU}{CPU}{central processing unit}
\newacronym{GPU}{GPU}{graphics processing unit}
\newacronym{QPU}{QPU}{quantum processing unit}
\newacronym{QPUs}{QPUs}{quantum processing units}
\newacronym{HPC}{HPC}{high performance computing}
\newacronym{LRZ}{LRZ}{Leibniz Supercomputing Centre}
\newacronym{DSL}{DSL}{domain-specific language}
\newacronym{eDSL}{eDSL}{embedded domain-specific language}
\newacronym{NISQ}{NISQ}{noisy intermediate-scale quantum}
\newacronym{MPI}{MPI}{Message Passing Interface}
\newacronym{QRAM}{QRAM}{quantum random access machine}
\newacronym{HQCC}{HQCC}{heterogeneous quantum-classical computation}
\newacronym{API}{API}{application programming interface}
\newacronym{QPT}{QPT}{quantum programming tool}
\newacronym{QC}{QC}{quantum computing}
\newacronym{ISA}{ISA}{instruction set architecture}
\newacronym{RISC}{RISC}{reduced instruction set computer}
\newacronym{RAM}{RAM}{random access memory}
\newacronym{IR}{IR}{intermediate representation}
\newacronym{SQRAM}{SQRAM}{sequential QRAM}
\newacronym{HPCQC}{HPCQC}{high performance computing - quantum computing}
\newacronym{QIR}{QIR}{quantum intermediate representation}
\newacronym{SQIR}{SQIR}{standardized quantum intermediate representation}
\newacronym{AOT}{AOT}{ahead-of-time}
\newacronym{JIT}{JIT}{just-in-time}
\newacronym{SW}{SW}{software}
\newacronym{HW}{HW}{hardware}
\newacronym{UQP}{UQP}{unified quantum platform}
\newacronym{QDMI}{QDMI}{quantum device management interface}
\newacronym{MQSS}{MQSS}{Munich quantum software stack}
\newacronym{QCP}{QCP}{quantum control processor}
\newacronym{QCPs}{QCPs}{quantum control processors}
\newacronym{QISA}{QISA}{quantum instruction set architecture}
\glsdisablehyper 

\usepackage{tikz}

\usepackage[colorinlistoftodos,prependcaption,textsize=tiny]{todonotes} 
\usepackage{float}

\usepackage{multirow}
\usepackage{subfigure}

\newcommand\copyrighttext{%
  \footnotesize  © 20xx IEEE. Personal use of this material is permitted. Permission from IEEE must be
obtained for all other uses, in any current or future media, including
reprinting/republishing this material for advertising or promotional purposes, creating new
collective works, for resale or redistribution to servers or lists, or reuse of any copyrighted
component of this work in other works.}
\newcommand\copyrightnotice{%
\begin{tikzpicture}[remember picture,overlay]
\node[anchor=south,yshift=10pt] at (current page.south) {\fbox{\parbox{\dimexpr\textwidth-\fboxsep-\fboxrule\relax}{\copyrighttext}}};
\end{tikzpicture}%
}

\makeatletter
\newcommand{\linebreakand}{%
  \end{@IEEEauthorhalign}
  \hfill\mbox{}\par
  \mbox{}\hfill\begin{@IEEEauthorhalign}
}
\makeatother

\begin{document}
\title{Integration of Quantum Accelerators into HPC: Toward a Unified Quantum Platform
}

\author{
\IEEEauthorblockN{Amr Elsharkawy}
\IEEEauthorblockA{\emph{Chair of Computer Architecture and Parallel Systems} \\
\emph{Technical University of Munich}\\
Munich, Germany \\
amr.elsharkawy@in.tum.de}
\and
\IEEEauthorblockN{Xiaorang Guo}
\IEEEauthorblockA{\emph{Chair of Computer Architecture and Parallel Systems} \\
\emph{Technical University of Munich}\\
Munich, Germany \\
xiaorang.guo@tum.de}
\and
\linebreakand
\IEEEauthorblockN{Martin Schulz}
\IEEEauthorblockA{\emph{Chair of Computer Architecture and Parallel Systems} \\
\emph{Technical University of Munich}\\
Munich, Germany \\
schulzm@in.tum.de}
}

\maketitle

\begin{abstract}
To harness the power of \gls{QC} in the near future, tight and efficient integration of \gls{QC} with \gls{HPC} infrastructure (both on the \gls{SW} and the \gls{HW} level) is crucial. 
This paper addresses the development of a \gls{UQP} and how it is being integrated into the \gls{HPC} ecosystem. It builds on the concepts of hybrid \gls{HPCQC} workflows and a unified \gls{HPCQC} toolchain, introduced in our previous work and makes the next needed step: it unifies the low-level interface between the existing classical \gls{HPC} systems and the emerging quantum hardware technologies, including but not limited to machines based on superconducting qubits, neutral atoms or trapped ions. 
The \gls{UQP} consists of three core components: a \emph{runtime library,} an \emph{\gls{ISA}} and a \emph{\gls{QCP}} micro-architecture. In particular, this work contributes a unified \gls{HPCQC} runtime library that bridges the gap between programming systems built on \gls{QIR} standard with a novel, unified hybrid \gls{ISA}.
It then introduces the initial extension of an \gls{ISA} and \gls{QCP} micro-architecture to be platform and technology agnostic and enables it as an efficient execution platform.
The \gls{UQP} has been verified to ensure correctness. Further, our performance analysis shows that the execution time and memory requirements of the runtime library scale super-linearly with number of qubits, which is critical to support scalability efforts in \gls{QC} hardware.

\glsresetall
\end{abstract}

\copyrightnotice{}
\section{Motivation}\label{sec:Introduction}

The \gls{QC} paradigm offers the opportunity to tackle involved scientific problems, such as quantum chemistry simulations~\cite{chemical2020} and cryptography~\cite{securecom2020}, which have been considered intractable in the classical computing sense. 
The research community has recently realized that tangible quantum advantage is not going to be achieved in a vacuum, 
but requires integrating quantum accelerators into the currently available \gls{HPC} systems and workflows~\cite{ruefenacht2022ea}.
As a consequence, novel hybrid classical-quantum systems are emerging to push the boundaries of both fields
by attempting to tackle the
question of \emph{``what is the optimal way of integrating quantum and classical computations on the software and the hardware level''?}

One of the core approaches to achieve quantum advantage during the \gls{NISQ} era is via hybrid variational quantum-classical algorithms, which utilize both classical and quantum computing resources. 
These algorithms involve a substantial amount of communication between quantum and classical processors. 
In case of slow communication and high latency connection, such interaction will add a huge overhead that would eat away any advantage we gain from quantum acceleration. It, therefore, requires an efficient abstraction layer that unifies the interaction between classical and quantum machines and supports a seamless integration process.
In our previous workshop paper~\cite{Seitz_HPCQC}, we introduce the concept of a unified hybrid HPCQC toolchain that can dampen the overhead of communication between both sides, in order to maximize the benefit of the hybrid compute resources. In this work, we take this initial idea and provide new contributions that enable the low-level integration and interfacing with the quantum accelerator and how it is realized with currently available software tools and hardware components. 

To enable low-level integration, considerable efforts are currently underway to develop \gls{QCPs} as close as possible to \gls{QPUs}~\cite{Guo2023:iccd,fu_eqasm_2019,zhang2021exploiting,stefanazzi2022qick}. While these specialized controllers promise to deliver fast and timing-precise control to mitigate the impact of short decoherence times, the absence of unified \gls{QISA} support means each currently existing \gls{QCP} is limited to serving a single physical technology.  Expanding this to mutliple technologies is, therefore, critical, yet challenging to achieve.

To address these challenges, we propose a \gls{UQP} that abstracts away the complexities of integrating multiple quantum modalities with classical \gls{HPC} system. The \gls{UQP} considers both software and hardware perspectives. 
Overall, the main contribution of this work is as follows:
\begin{itemize}
  \item On the software level, the \gls{UQP} introduces a runtime library that maps a \gls{QIR} representation of the quantum kernel to the novel unified binary instructions understood by the \gls{QCP}. Moreover, the novel \gls{ISA} is capable of accommodating all instructions exposed by the various quantum hardware to the software side. 
  
  \item At the backend interface of the \gls{UQP}, we upgrade the \gls{QCP} based on our previous work \textit{HiSEP-Q}\cite{Guo2023:iccd}, by generalizing the architecture to also support neutral atom quantum computers. The upgraded version implements the quantum and classical control logic needed for not only initiating execution for distinct quantum hardware technologies but also post-processing the measurement results. 

  \item Our experimental evaluations illustrate that the execution time and memory utilization of the runtime library scales super-linearly with the number of qubits. This feature guarantees the applicability of our work in large-scale quantum systems, existing and future.  
\end{itemize}


The remainder of the paper is structured as follows. 
In \Cref{sec:Background}, we introduce the core concepts of \gls{HPCQC} integration and workflow, as well as available quantum \gls{HW} technologies and their interface to the software stack. 
In \Cref{sec:RelatedWork} we highlight the relevant ongoing research and the shortcomings of the currently available workflows. 
In \Cref{sec:UQB} we present our proposal for the \gls{HPCQC} workflow of a \gls{UQP} and
in \Cref{sec:eval} we validate the workflow and analyze its performance properties.
Finally, the work concludes with \Cref{sec:Conclusion}.

\section{Background}\label{sec:Background}
\begin{figure}[bt]
    \centering
    \includegraphics[scale=1.8]{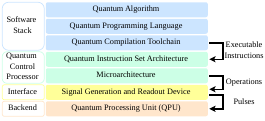}
    \caption{Abstract representation of a full-stack quantum computer.}
    \label{fig:abstractfullstack}
\end{figure}

\begin{figure}[bt]
    \centering
    \includegraphics[scale=1]{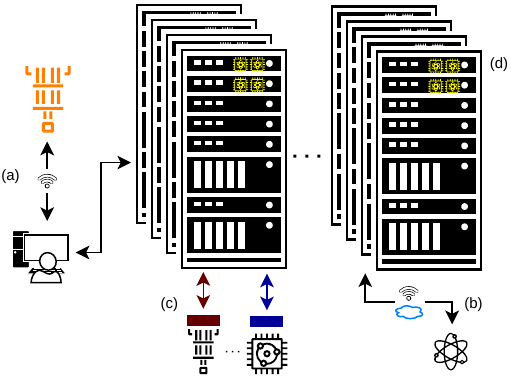}
    \caption{\gls{HPCQC} Integration Scenarios. (a) Loose Integration -- Standalone, (b) Loose Integration -- Co-located, (c) Tight Integration -- Co-located and (d) Tight Integration -- On-node. }
    \label{fig:HWintegration}
\end{figure}

\gls{HPCQC} is a comprehensive undertaking that must span many layers, starting from the hybrid algorithmic description, all the way to the interface between classical and quantum \glspl{HW}, see \Cref{fig:abstractfullstack}.
In this section, we describe the concepts needed for \gls{HPCQC} integration from different perspectives.
\subsection{User View on HPCQC Integration} \label{subsec:HPCQCInteg}
While \gls{QC} offers distinct benefits over classical computing in particular contexts, its utility is constrained by the presently available qubit count in quantum \glspl{HW} and the absence of effective methods for storing, manipulating, and retrieving classical data within quantum frameworks. Consequently, the prevalent approach to leveraging \gls{QC} is to deploy a \gls{QPU} as a supplementary, sophisticated accelerator that can be used for specific computational tasks~\cite{ruefenacht2022ea}.
\subsubsection{SW-Level View on HPCQC Integration} \label{sssec:SWInteg}
\gls{HPCQC} integration, from the \gls{SW} perspective, entails the design and development of programming models, execution schedulers, runtime systems, and networking approaches that not only take into consideration the core features of quantum computation, e.g., stochastic behavior and real-time feedback control, but also integrate well with existing and emerging classical computing approaches~\cite{Elsharkawy_integration,travis_sw}. The ultimate goal of the \gls{SW} tools built to realize all these functionalities, is to allow for efficient and seamless integration that does not eat away the expected quantum \gls{HW} advantage by overheads. 

\subsubsection{HW-Level View on HPCQC Integration} \label{sssec:HWInteg}
\gls{HPCQC} integration, from the \gls{HW} perspective, focuses on the physical arrangement of quantum \gls{HW} and \gls{HPC} infrastructure. 
Specifically, it considers the architecture of \gls{HPC} and quantum \gls{HW} elements and their interconnections~\cite{humbleQuantumComputers2021}.   
In our previous work~\cite{Elsharkawy_integration, Elsharkawy_challenges}, we have identified four integration scenarios that could co-exist as well as the technical challenges they introduce, see~\Cref{fig:HWintegration}. The different scenarios represent the progress of the emerging field of \gls{HPCQC} integration, starting from standalone \glspl{QPU} to on-node integration, which comes with the highest technical challenges.

In the \gls{NISQ} era, \gls{HW}  vendors are investigating different technologies (e.g., superconducting materials, atoms \& ions) to build a quantum \gls{HW}. Each of these has unique characteristics and imposes different constraints. This fact not only complicates the task of supercomputing centers that work on integrating quantum devices with their \gls{HPC} systems, but also raises the bar very high for researchers who work on developing software tools that interface with such \glspl{HW}. 
\subsection{The HPCQC Workflow} \label{subsec:HPCQCWorkflow}

\begin{figure}[tb]
    \centering
    \includegraphics[scale=1.7]{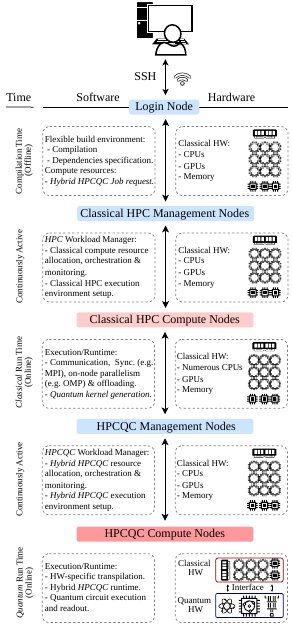}
    \caption{Abstract HPCQC workflow. The figure shows five components: login node, HPC management node, HPC compute node, HPCQC management node and HPCQC compute node.}
    \label{fig:abstractworkflow}
\end{figure}

\Cref{fig:abstractworkflow} shows an abstract representation of a hybrid \gls{HPCQC} workflow. 
The workflow describes a sequence of steps to perform a computation on a hybrid \gls{HPCQC} system.
The description highlights the location and the time of the tasks being executed. 
The diagram is broadly divided into three main components: first are the login and management nodes, second are the classical compute resources, and third are the hybrid \gls{HPCQC} compute resources and the quantum hardware. 
\Gls{HPC} systems usually feature an external access point, often known as a login node. 
This node offers a versatile building environment, equipped with a package manager, enabling users to establish their necessary dependencies and choose their desired compiler or interpreter. 
The compiled executables and the specified resources are encapsulated within a job request, which is submitted to a workload manager. 
This marks the end of the \emph{compilation time}.

During \emph{classical run time}, the program execution starts simultaneously on the allocated classical compute nodes, where communication and synchronization occur according to the user's specifications within the program with the help of \gls{HPC} libraries and frameworks, such as \gls{MPI} and OpenMP~\cite{OMP, mpi40}. 
During this time, it is typical that all information (e.g., run time parameters) needed for quantum circuit generation is available. 
Hence, on the classical \gls{HPC} nodes, quantum kernels are generated, and offloaded to the \gls{HPCQC} resource manager, which then allocates the suitable resources and initiates execution which marks the start of \emph{quantum run time}. 

The details of how quantum and classical resources interplay within \gls{HPCQC} compute resources during \emph{quantum run time} is an open research question.
This is because it is highly dependent on the involved classical systems, quantum systems, available level of interfacing, and how much control is exposed to the programmer (i.e., which execution model is adopted?~\cite{Elsharkawy_integration}).
In the process, though, the quantum circuit is optimized according to the allocated quantum hardware and offloaded to the \gls{QPU} to get executed.
Upon completion, the system provides the user with the job outcomes, including both classical and quantum results.

\subsection{Quantum Intermediate Representation Specification (QIR)} \label{subsec:QIR}
\Gls{QIR}~\cite{qir_specification} is a specification that defines an intermediate representation for quantum programs, designed to bridge the gap between high-level quantum programming languages and lower-level quantum execution targets, such as quantum hardware or simulators. 
\Gls{QIR} is built upon the well-established LLVM IR, leveraging its robust, platform-agnostic framework to enable the optimization and execution of quantum programs. 
The goal of \gls{QIR} is to provide a common interface that facilitates interoperability between different quantum programming languages and diverse quantum processing units (QPUs).

The \gls{QIR} specification is actively developed and maintained by the \gls{QIR} Alliance, a collaboration of industry, academia, and independent professionals. 
The specification is still evolving, with ongoing efforts to expand its capabilities, improve its efficiency, and enhance its compatibility with a wide range of quantum hardware and classical integration scenarios. 
The open-source nature of the project encourages contributions from the \emph{HPCQC} community, fostering innovation and the adoption of best practices in software development.

\subsection{Quantum Hardware and Control} \label{subsec:QHW}
To achieve the full potential of quantum advantages, SW-Level and HW-Level should be seamlessly integrated, which also contributes to the fully programmable quantum computer shown in Figure~\ref{fig:abstractfullstack}. Although the high-level programming language can be compiled into hardware-specific instructions, direct control from the SW level to the HW level restricts the efficiency and scalability of the quantum computers\cite{fu2017experimental}. Therefore, \gls{QCPs}\cite{Guo2023:iccd,Guo2023:fpl} are proposed to act as an interface between the software and hardware layers. QCPs compile the executable binary instructions generated by compilers into a sequence of pulses to control the qubits. Positioned close to \gls{QPUs}, this kind of unit is able to provide precise nanosecond timing control and features fast mid-circuit measurement support. 
\section{Related Work and Research Gaps}\label{sec:RelatedWork}

It is evident that existing literature predominantly concentrates on the development of hybrid programming models and the seamless integration of quantum software stacks within conventional HPC environments. 
On the one hand, the authors of~\cite{mintz_qcor_2019, mccaskey_extending_2021, nguyen2022extending, mccaskey_xacc_2020} have proposed language extension (i.e. C++ and Python) to integrate quantum and classical computations. Moreover, they have proposed a system-level software infrastructure that supports such extensions. 
These studies explore frameworks that allow quantum and classical computing resources to collaboratively address computational tasks, leveraging the strengths of each to enhance overall performance and efficiency.
These efforts correspond to the software stack components of~\Cref{fig:abstractfullstack}.
On the other side, data centers that host classical and quantum resources are investigating the optimal approach to build a unified software stack that supports the emerging programming models. For instance, the \gls{LRZ} has proposed a vision on how the two paradigms may come together efficiently~\cite{ruefenacht2022ea, Schulz2022AHQ}. Moreover, there have been other efforts investigating hybrid HPCQC hardware architectures~\cite{humbleQuantumComputers2021} and middleware setup~\cite{Luckow_middleware}. 

Despite these advancements, a noticeable gap in the research landscape is not only the \emph{uniformity} of binary instructions across different quantum modalities, but also the software infrastructure needed to map higher-level representations to such unified binary representation. 
Current discussions largely bypass the challenges of creating a standardized set of binary instructions that would ensure \emph{compatibility} and \emph{interoperability} among varied quantum computing platforms. 
Such standardization would allow quantum applications, once compiled, to be executed across different quantum systems without the need for recompilation or modification, addressing a crucial bottleneck in the widespread adoption of quantum accelerators in HPC settings. 

Additionally, there also exists a gap in the development of novel quantum control processors capable of handling the aforementioned standardized binary streams and providing cross-technology control, see~\Cref{fig:abstractfullstack}. 
To date, most existing quantum controllers~\cite{fu_eqasm_2019,qubic2,stefanazzi2022qick,Xu2021,Mathews2022} are all designed for a single physical modality. To the best of our knowledge, OpenQL~\cite{Openql}, as proposed by Khammassi~et~al., stands as the sole example of an architecture capable of controlling both superconducting and semiconducting qubits. However, this design relies on the eQASM instruction set~\cite{fu_eqasm_2019}, which is topology-dependent and lacks scalability. Consequently, the utilization of OpenQL is limited in scenarios where qubit connectivity is unconstrained, such as in neutral atom quantum computers. Therefore, addressing these gaps could significantly streamline the deployment of quantum solutions, fostering a more integrated and versatile \gls{HPCQC} landscape.


\section{Introducing a Unified Quantum Platform}
\label{sec:UQB}

The concept of a unified quantum platform is an emerging \gls{HPCQC} system that consists of various building blocks including various \gls{SW} tools and \gls{HW} components.
In the following, we highlight our focus, which is the interfacing layer between the hybrid \gls{SW} toolchain and the quantum processing units. 
First, we describe our novel unified runtime library implementation. 
Then, we discuss the extension of a single technology approach and introduce a generalized \gls{ISA} and \gls{QCP} micro-architecture as a step toward supporting more backend modalities.


\subsection{Unified HPCQC Runtime Environment} 
\label{subsec:ccompilation}

In the \gls{NISQ} era, no specific technology has proved itself as the superior approach to building quantum hardware. 
Hence, supercomputer centers are poised to have to host various \glspl{QPU} based on different technologies. 

Looking at the evolution 
\Cref{fig:crossTechnology} shows the integration development of distinct quantum accelerators into currently available HPC systems and workflows.
(a) At the early stage of the emerging HPCQC field, each backend requires a standalone toolchain to map the algorithmic representation to a unique low-level representation that is understandable by the machine, 
(b) then, the concept of a unified toolchain has emerged, yet, it still requires providing different interfacing approaches that suit the available hardware, 
(c) then various efforts, including the one within the \gls{LRZ}, push toward a unified intermediate representation interface where all backends accept the same languages, such as QASM and QIR, but still leave us with distinct technology stacks towards the hardware. 
To really provide a cross-technology platform, in this work, we propose to go to the final step,
(d), which pushes the unified interface as low as the \gls{ISA} level. 
For this, we exploit two main building blocks, developed or adopted in this work for our \gls{HPCQC} approach:


\begin{figure}[h!]
    \centering
    \includegraphics[scale=1.393]{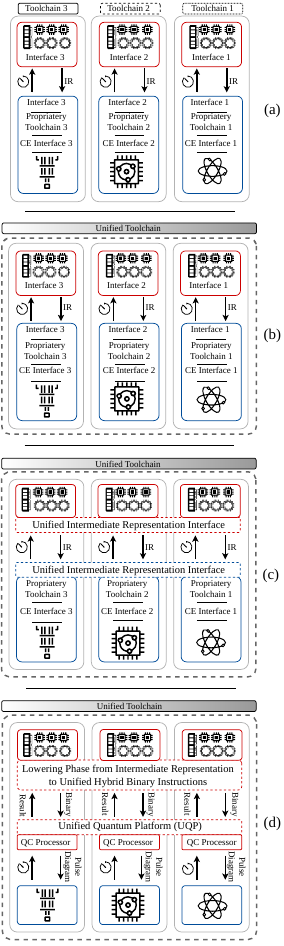}
    \caption{The expected progress of cross-technology execution within an \gls{HPCQC} system.}
    \label{fig:crossTechnology}
    \vspace{-5mm}
\end{figure}

\subsubsection{Block 1: Unified Intermediate Representation}\label{sssec:UQIR}
We first must realize a truly cross-platform intermediate representation. 
For this, we adopt the \gls{QIR} specification as an established, platform- and vendor-neutral approach.
\gls{QIR} does not enforce a precise syntax; it is the responsibility of quantum hardware builders and software developers to adopt the syntax that they see fit for their platform.
As an initial step, we have adopted the \emph{base profile} that only includes quantum operations and limited classical instructions. 
In future work, we will incorporate new instructions as they emerge from both sides of the hardware builders and the software developers.


\subsubsection{Block 2: Unified Runtime Library} \label{sssec:urt}
Building on \gls{QIR}, we design and implement a unified runtime library that maps the \gls{QIR} base profile representation to the novel unified \gls{ISA} supported by the proposed quantum control processor, which is described below.
The library is written in C\textbackslash{}C++.
The runtime takes care of implementing the binary instruction corresponding to the \gls{QIR} representation along with the required memory management for allocating quantum and classical registers.
At the moment we only support sequential execution of the 32-bit binary instructions. 
The library implementation addresses the time management requirements for instruction scheduling.
Moreover, the runtime library is responsible for initializing the execution environment which differs according to the targeted quantum hardware. 
Yet, across all hardware modalities, the execution environment initialization retrieves information such as the size of both quantum and classical registers from the attributes of the QIR representation. 
On a different note, additional information, such as the number of shots and the specific hardware target, is provided to the runtime via the workload manager as part of the job submission call. 
As both the input (i.e., represented by the \gls{QIR} specification) and the output (i.e., represented by the novel \gls{ISA}) evolve, the runtime implementation is going to continuously be updated to support the latest setup.   

\subsection{SW-HW Interface} 
\label{sssec:swinterface}
The binary file generated by the runtime system, including both classic control and quantum instructions, is offloaded to the quantum control processor via a shared memory segment. 
The quantum control processor has an on-board ARM CPU that takes care of making the binary available to the on-board controllers, which are then used to execute the quantum circuit. While this interface is simple, it is both effective and efficient, allowing quick and easy transfer of quantum code to hardware execution.


\subsection{Novel Quantum Control Processor} \label{subsec:QCP}

\begin{figure}[bt]
    \centering
    \includegraphics[scale=1.7]{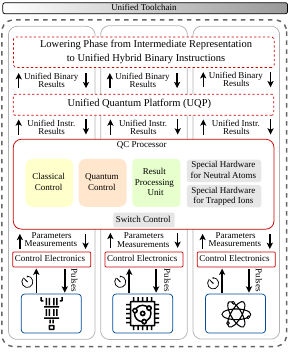}
    \caption{System architecture of the proposed cross-technology quantum control processor.}
    \label{fig:CT-QCP}
\end{figure}

While the runtime system, with its intermediate representation and its execution platform, can hide the already discussed differences between physical modalities -- from superconducting qubits to neutral atoms and ion traps -- the actual technology stack on the backend must also support this heterogeneity. 
However, even though the physical technologies of the platforms are different, the control theories over them have much in common. 
This provides us with the needed leverage to create a technology-agnostic backend architecture, which we propose in our novel system architecture for a unified control unit, illustrated in Figure~\ref{fig:CT-QCP}. 
When selecting a physical platform to execute the algorithm, the unified toolchain compiles high-level instructions into executable unified binary instructions. Meanwhile, the switch also selects the target modality. 
After loading the waveform configuration files for each platform, the generated executable binary sequence will then be processed on the QCP to generate the pulses (through control electronics) to control the qubits, specialized for the particular modality as defined by a respective pulse library. 
The pulse library itself is defined by the backend developer and, with that, ensures the needed flexibility while hiding the differences in the user-facing interface. 
Further, to improve efficiency, the architecture can be enhanced with specific accelerators, e.g., for handling of atoms or swap optimizations in ions, that can be used on demand. 
This concept of heterogeneous accelerators is well known from the development of modern multi-media and mobile processors and has been shown to be highly performing and energy efficient.

To implement this HW-Level integration, we again build on two main building blocks: a novel cross-technology control processor and a new instruction set that supports arbitrary, pulse-based technology platforms.

\subsubsection{Quantum Instruction Set} \label{sssec:QISA}
A \gls{QISA} functions as an interface between the compilation toolchain and the control microarchitecture. Its specifications have a significant impact on both the performance and scalability of QCPs, as well as the efficiency of high-level algorithm mapping. Given the current development state of HPCQC, where quantum computers serve as specialized accelerators alongside classical HPC systems, it becomes essential to design \gls{QISA} in a way that is compatible with classical ISAs. Furthermore, to effectively manage classical control flows, such as conditional jumps and feedback management, \gls{QISA} needs to be hybrid in nature. This implies that classical and quantum instructions need to seamlessly integrate with each other. 

\begin{figure}[bt]
    \centering
    \includegraphics[scale=2]{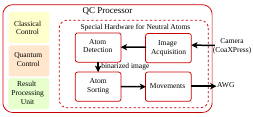}
    \caption{Architecture of the QCP activated for neutral atom quantum computer. AWG represents an arbitrary waveform generator.}
    \label{fig:new_ma}
\end{figure}

In order to achieve this goal, we build on a \gls{QISA} for superconducting systems~\cite{Guo2023:iccd}, which features a hybrid instruction format and an efficient and scalable control for that modality
To achieve a unified control scheme, we extend the support in this \gls{QISA} beyond superconducting qubits alone.
We do this by adding instructions needed on other platforms, as well as their needed support functionality, like the introduction of swap gates. In our current work, we demonstrate this by supporting both superconducting and neutral atom-based systems in a single platform and \gls{QISA}.

While the fundamentals of qubit addressing and gate control remain unified, with only waveform parameters differing across platforms, we have enriched the instruction set to incorporate actions specific to neutral-atom systems.
This enrichment mainly involves introducing instructions for atom preparation, such as image acquisition, atom detection, and atom resorting, which facilitate the seamless integration of neutral-atom quantum computing into the unified \gls{QISA} framework. Moreover, to streamline compiler efforts and maintain consistency across platforms, these initialization instructions are strategically placed at the beginning when configuring the neutral-atom quantum computer as the target platform. Table~\ref{tab:inst} provides an illustration of the extended instructions, while the original \gls{QISA} specification can be found in the original Hisep-Q processor~\cite{Guo2023:iccd}.

\begin{table}[h]
\centering
\caption{Extended Instructions for Neutral-Atom Quantum Computer}
\begin{tabular}{|c|c|}
\hline
\textbf{Function} & \textbf{Description} \\
\hline
\hline
Image Fetch &  Start to fetch the atom image into the memory. \\
\hline
Atom Detection  & Start to detect the atom positions and occupancy \\
\hline
Atom Sorting  & Start to sort (rearrange) atoms to a defect-free target\\
\hline
Atom Moving  & Start to send control signals  \\
\hline
\end{tabular}
\label{tab:inst}
\end{table}

\subsubsection{Microarchitecture}\label{sssec:MArch}
Along with the changes we have made on the \gls{QISA}, we also enhance the microarchitecture accordingly. Figure~\ref{fig:CT-QCP} demonstrates the unified microarchitecture targeting cross-technology support, by adding technology technology-specific accelerator block; a corresponding functional block will be activated when a physical platform is selected, a technique readily used in the mobile and multi-media chip world in which one processor offers a variety of small custom blocks for specific target functions. Figure~\ref{fig:new_ma} exemplifies the microarchitecture in managing a neutral atom quantum computer. While the three core blocks overseeing classical and quantum control, along with processing intermediate results, remain consistent, a specialized hardware configuration is introduced in this scenario. Corresponding to our extended instructions, the fluorescence image, captured by an electron-multiplying charge-coupled device (EMCCD) camera, is acquired within the 'Image Acquisition' stage~\cite{jonas2023}. Subsequently, an atom detection unit, typically employing deconvolution algorithms, detects the presence of atoms. Following this, an atom-resorting algorithm is engaged to devise a strategy for arranging the atoms into a defect-free atom array. Finally, this strategy is translated into a sequence of commands instructing arbitrary waveform generators (AWGs) to manipulate the qubits efficiently. By activating and deactivating specialized hardware, our quantum controller can be switched between different platforms effectively while maintaining the same basic unified components.

\section{Experiment and Evaluation} \label{sec:eval}

To show the correctness and the efficiency of the presented approach, we evaluate our unified system with using a hybrid workflow. First, we conduct a performance analysis to show that our implementation does not add significant run time overhead. We then validate the correctness of the generated binary instructions step-by-step to guarantee the correspondence to the high-level algorithmic description. 

\subsection{Performance Analysis} \label{sec:analysis}

\begin{figure}[htp]
  \centering
  \subfigure[Memory resources scalability.]{\includegraphics[scale=0.4]{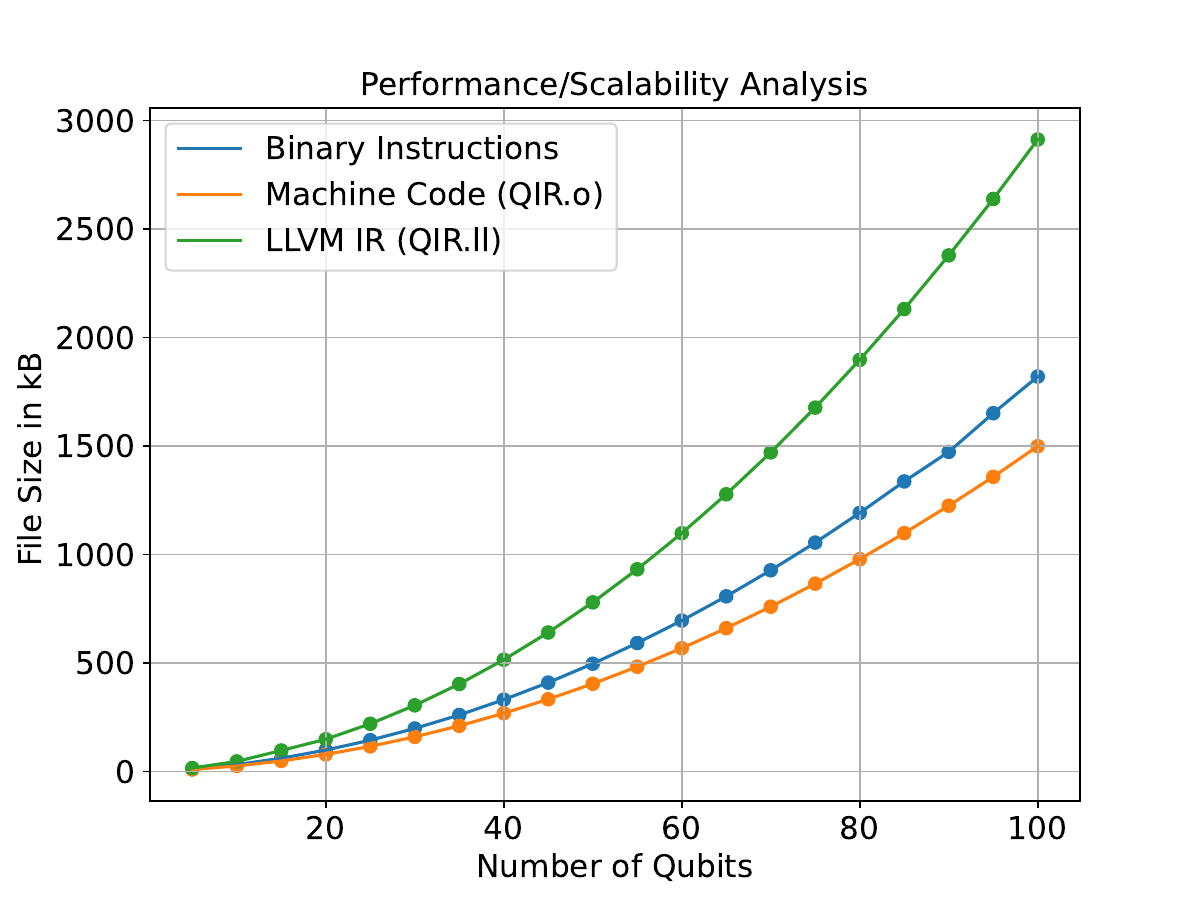}\label{fig:pfa}}
  
  \subfigure[Execution time scalability.]{\includegraphics[scale=0.4]{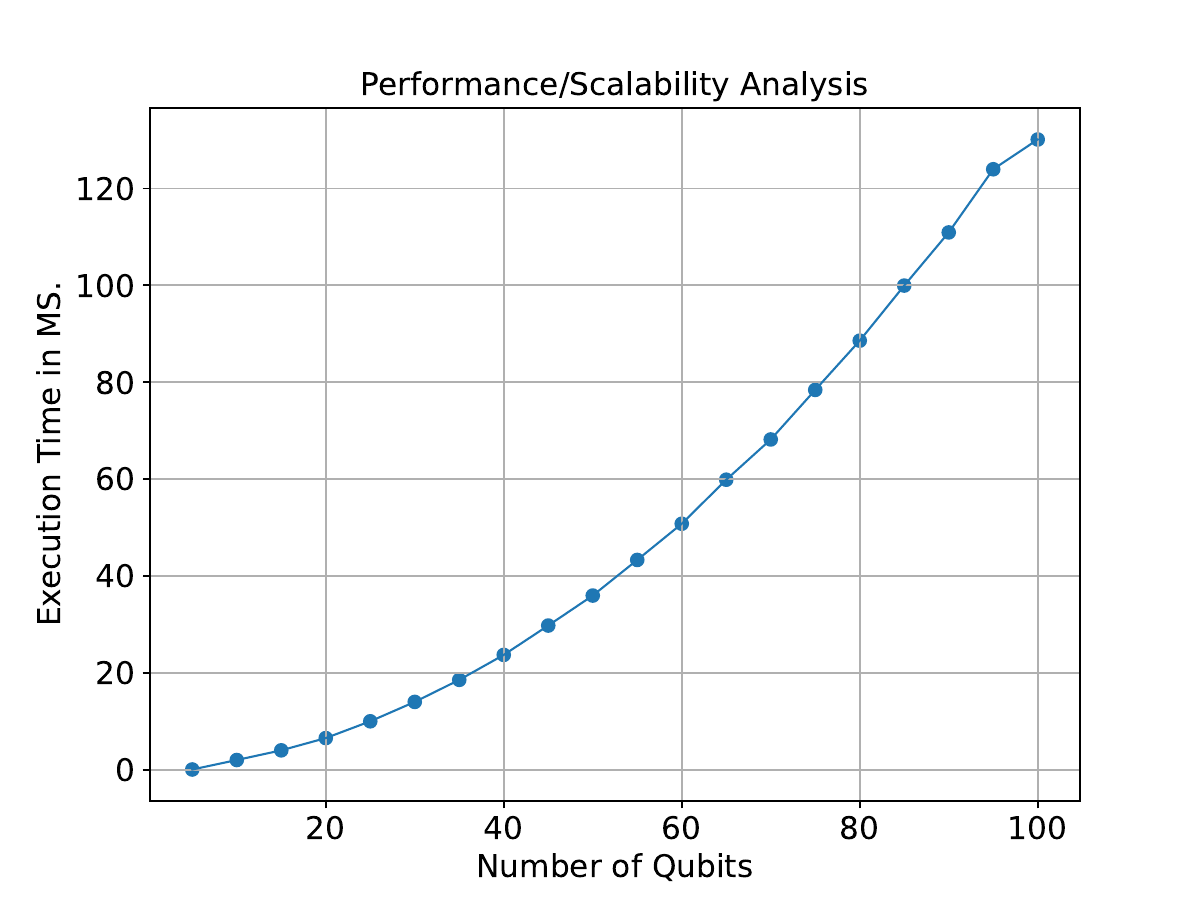}\label{fig:pfb}}
  \caption{Performance analysis of the runtime library - Alpha version.}
  \label{fig:performanceanalysis}
\end{figure}

In the experiment, we utilize the \emph{Munich Quantum Toolkit (MQT) Bench}~\cite{Quetschlich_2023}, which is a benchmarking software and design automation tool for quantum computing.
The analysis has been conducted on a local machine that runs Linux as the operating system with a 13th Gen Intel(R) Core(TM) i9-13900HX CPU. The study scales the number of qubits from five to one hundred, which is the maximum number of qubits that can be addressed by the current version of the 32-bit \gls{QISA}. The result sample in~\Cref{fig:performanceanalysis} is for the Amplitude Estimation algorithm represented in the native gate set of IBM quantum hardware. Each data point is the average of a thousand executions. 
\Cref{fig:pfa} shows that the memory requirement of our implementation scales super-linearly with the size of the input circuit. 
Hence, there is no exponential increase in memory resources despite of introducing the memory and time management instructions on the binary level representation. 
Moreover, \Cref{fig:pfb} illustrates that the execution time of the runtime library also scales super-linearly with the size of quantum code. 
The super-linear scaling of memory and execution time facilitates predictable resource planning and allocation. This predictability is crucial for scaling quantum applications as it allows developers and system architects to estimate the computational and memory requirements based on the number of qubits involved easily.
Such performance means that -- as quantum computers grow in terms of qubit count -- the software infrastructure can expand to accommodate this growth without facing exponential increases in resource demands. This scalability is essential for the practical implementation of larger and more complex quantum algorithms.
This provides the basis for future work that includes more run time analysis and optimization.

\subsection{Workflow Verification} \label{sec:workflowverification}

\begin{figure*}[tb]
    \centering
    \includegraphics[scale=1.15]{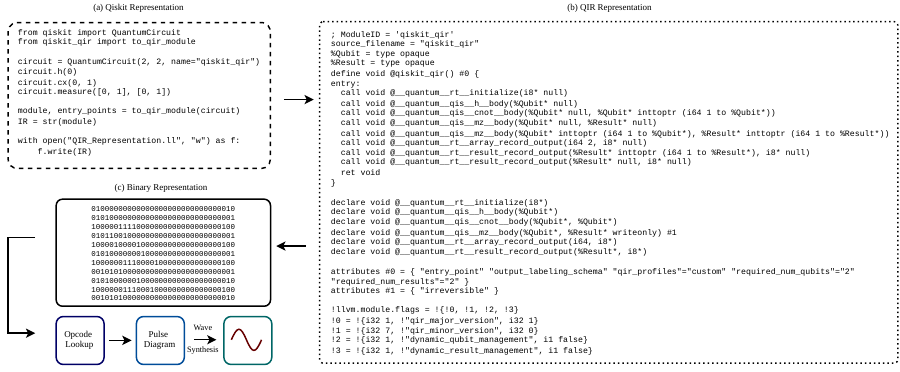}
    \caption{Workflow verification.}
    \label{fig:workflowverification}
\end{figure*}

The work presented adds the missing building block in existing \gls{HPCQC} workflows, as the one being built by LRZ, as it bridges the gap between the \gls{QIR} representation of the quantum kernel and the novel \gls{ISA} supported by the unified quantum platform. 
This mapping enables researchers, for the first time, to target the \gls{UQP} using the high-level representation of quantum circuits without the need for writing binary instructions manually.
Moreover, our implementation lays the groundwork for a more comprehensive, more intricate \gls{UQP} which abstracts away the complexity of interfacing with different quantum hardware modalities. 
This research, from the software perspective, introduces the concept of binary portability across different quantum machines as long as requirements such as qubit connectivity are satisfied. 
Hence, there is no need to re-compile to generate a new binary to target different quantum processing units. 

\Cref{fig:workflowverification} shows the workflow verification of the proposed setup. 
The high-level quantum circuit representation is mapped to a quantum intermediate representation.
\Cref{fig:workflowverification}(a) illustrates a simple quantum circuit that prepares the famous bell state using Qiskit as a development environment. 
The mapping from  Qiskit representation to \gls{QIR} representation is done via a software tool written by Microsoft developers called \texttt{"qiskit\_qir"}.
The corresponding \gls{QIR} implementation is seen in \Cref{fig:workflowverification}(b), where each quantum or classical instruction is represented by an external function call to the backend runtime library. 
The runtime library is responsible for the specifics of implementing such instruction to fit the supported hardware. 
Moreover, additional metadata is included in the \gls{QIR} representation, such as the attributes.  
Then, upon compilation and execution of the \gls{QIR} code, the corresponding binary instructions are generated and offloaded to the unified quantum platform as seen in \Cref{fig:workflowverification}(c).
Finally, the control logic implemented on the \gls{QCPs} generates the parameters required for the wave synthesis which is then passed to the control electronics for the actual quantum operation execution on the quantum register.

\Cref{tab:verification} shows the direct correspondence between the high-level representation of the simple Bell state quantum circuit and the binary instructions being executed on the quantum control processor. For the sake of clarity, this example intentionally avoids any intermediate optimization passes that would change the structure of the quantum circuit. This correspondence table is indicative of a well-defined ISA for quantum computing. It reflects the design choices made for the quantum processor's ISA, such as operation types, operand specifications, and memory access patterns. Moreover, The table suggests a systematic approach to extending the binary instruction set to include new quantum operations as the field of quantum algorithms grows. 

\begin{table*}[h]
\centering
\caption{Binary Instruction Verification}
\begin{tabular}{|l|l|}
\hline
 \multicolumn{1}{|c|}{\textbf{Qiskit Representation}} & \multicolumn{1}{|c|}{\textbf{Binary Instruction Representation} }\\
\hline
\hline
 N/A & 01000000000000000000000000000010 - Execution environment initialization \\
\hline
\multirow{2}{12em}{circuit.h(0)} & 01010000000000000000000000000001 - Memory instruction \\
                                 & 10000011110000000000000000000100 - Hadamard operation \\
\hline
\multirow{2}{12em}{circuit.cx(0, 1)}  & 01011001000000000000000000000001  - Memory instruction \\
                                      & 10000100001000000000000000000100  - CNOT operation \\
\hline
\multirow{6}{12em}{circuit.measure([0, 1], [0, 1])}  &  01010000000100000000000000000001 - Memory instruction \\
                                 &  10000001110000100000000000000100 - First qubit measurement operation \\
                                 &  00101010000000000000000000000001 - Fetch last measurement \\
                                 &  01010000001000000000000000000010 - Memory instruction \\
                                 &  10000001110001000000000000000100 - Second qubit measurement operation \\ 
                                 &  00101010000000000000000000000010 - Fetch last measurement \\
\hline
\end{tabular}
\label{tab:verification}
\end{table*}

By providing a low-level view of the binary instructions, developers and researchers gain an in-depth understanding of how their quantum algorithms are executed on the hardware. This level of transparency ensures that the quantum operations are carried out precisely as intended, without the abstractions or optimizations that vendor-supplied runtimes might perform unknowingly to the users. It empowers users to make informed decisions regarding their quantum code, facilitating fine-tuned optimizations and adjustments based on the detailed feedback from the execution layer.

\section{Conclusion} \label{sec:Conclusion}
This work introduced the implementation of an abstraction layer needed to realize a unified quantum platform. 
Such implementation is positioned right at the interface of the software stack and quantum processing units. 
Hence, the implementation touches on both levels of software and hardware.  
On the \emph{software} side, we have developed a novel unified runtime library that maps QIR representation to an extended, unified hybrid ISA.
On the \emph{hardware} side, we show how a quantum control processor can be extended to include the logic to support multiple technologies and we demonstrate this by adding the required control logic for a quantum hardware based on neutral atoms technology in addition to the existing superconducting support. 
The performance analysis and workflow verification highlight the scalability capabilities of this approach and ensure the correctness of the implementation. 
To the best of our knowledge, this research is the first to successfully introduce the idea of a unified and open quantum platform and to implement it within a tight HPCQC integration setup.


\section*{Acknowledgment}

The research is part of the Munich Quantum Valley (MQV), which is supported by the Bavarian state government with funds from the Hightech Agenda Bayern. Moreover,  this work is supported by BMW Group.

\bibliographystyle{IEEEtran}
\bibliography{IEEEabrv, IEEEexample, references} 
\end{document}